\documentstyle [12pt,epsf] {article}

\def\yr{\,{\rm yr}}

\begin{document}
\begin{titlepage}
\begin{center}
\vspace{2cm}
\LARGE
A Unified Model for the Evolution of Galaxies and Quasars
\\                                                     
\vspace{1cm} 
\large
Guinevere Kauffmann \& Martin Haehnelt \\
\vspace{0.5cm}
\small
{\em Max-Planck Institut f\"{u}r Astrophysik, D-85740 Garching, Germany} \\
\vspace{0.8cm}
\end{center}
\normalsize
\begin {abstract}
We incorporate a simple scheme for the growth of supermassive black
holes into semi-analytic models that follow the formation and evolution
of galaxies in a cold dark matter dominated  Universe. 
We assume that supermassive black holes are formed and fuelled 
during major mergers. If two galaxies of 
comparable mass merge, their central black holes coalesce and a 
few percent of the gas in the merger remnant is 
accreted by the new black hole over a  timescale of a few times $10^7$
years. With these simple assumptions, our model not only fits many  
aspects of the observed evolution of galaxies, but  also reproduces 
quantitatively the observed relation between bulge luminosity and black hole
mass  in nearby galaxies, the strong  evolution of the quasar population with redshift,
and the relation between the luminosities of nearby quasars and those of their host galaxies.   
The strong decline in the number density of quasars   
from $z \sim 2$ to $z=0$ is due to the combination of three effects:              
i) a decrease in the merging rate, ii) a decrease in the  amount of 
cold gas available to fuel
black holes, and iii) an increase in the timescale for gas accretion.
In a $\Lambda$CDM cosmology the predicted decline in the total content of cold
gas in galaxies is consistent with that inferred from
observations of damped Lyman-alpha systems. Our results strongly
suggest             
that the evolution of  supermassive black holes, quasars and starburst galaxies
is inextricably linked to the hierarchical build-up 
of galaxies.  
\end {abstract}
\vspace{1.3cm}
\end {titlepage}

\section {Introduction}          

Soon after quasars were discovered, it was suggested that they are 
powered by the accretion of gas onto supermassive black holes at 
the centres of galaxies (Lynden Bell 1969).
The space density of luminous quasars is, however, two 
order of magnitudes smaller than that of bright galaxies and  
there has been a long-standing debate whether quasars 
occur only in  a subset  of galaxies or whether  all galaxies harbour a quasar
for a short time (Rees 1984, Cavaliere \& Szalay 1986, Rees 1990).
In recent years there has been  mounting  observational evidence 
that the evolution of normal galaxies and quasars is closely 
linked and that quasars are short-lived.                                 
The evolution 
of the total star formation rate density  of the Universe, 
the space density of starbursting galaxies and that of luminous 
quasars appear to be remarkably similar.  All three show a strong 
increase of more than an order of magnitude from $z=0$ to $z \sim 2$  
(Boyle \& Terlevich 1998, Dickinson et al. 1998, Sanders \& Mirabel 1996).
There is also increasing dynamical evidence
that supermassive black holes reside at the centres of most galaxies 
with substantial spheroidal components. The masses of the black holes scale linearly with the
masses of the spheroids, with a constant of proportionality in the range $\sim 0.002-0.006$.
(Kormendy \& Richstone 1995;  Magorrian et al. 1998; 
van der Marel 1999). 
The mass function of nearby black holes derived using the observed bulge luminosity--black hole mass
relation is also consistent
with that inferred from the 5GHz radio emission  in galactic cores and from the quasar luminosity
function itself (Franceschini, Vercellone \& Fabian 1998; Salucci et al. 1999).    
These results  strongly support the idea that
QSO activity, the growth of supermassive black holes and the formation of spheroids are
all closely linked (e.g. Richstone et al. 1998; Haehnelt, Natarajan \& Rees 1998; 
Cattaneo, Haehnelt \& Rees 1999). 

The most striking observed property of quasars                           
is their strong evolution with redshift. A number of papers have shown
that the rise in the space density of bright quasars from the 
earliest epochs to a
peak at $z \sim 2$ can be naturally explained 
in hierarchical theories of structure formation 
if the formation  of black holes is linked to the collapse of the first 
dark matter haloes of galactic mass (e.g. Efstathiou \& Rees 1988, 
Carlberg 1990, Haehnelt \& Rees 1993, Cavaliere, Perri \& Vittorini
1997, Haiman \& Loeb 1998). None of these papers provided more than a 
qualitative explanation for why the present abundance of bright quasars 
is two orders of magnitude below that at $z \sim 2$ 
(see also Blandford \& Small 1992).

In this paper, we focus on the low redshift evolution of the quasar population 
and its connection to the hierarchical build-up of  
galaxies  predicted in cold dark matter (CDM) -type cosmologies.                  
The formation  and evolution of galaxies in such cosmologies 
has been studied extensively
using {\em semi-analytic} models of galaxy formation.
These models follow the formation and evolution of galaxies within a merging hierarchy
of dark matter halos. Simple prescriptions are adopted to describe gas cooling, star formation,
supernova feedback and merging rates of galaxies. Stellar population
synthesis models are used to generate galaxy luminosity functions, counts and redshift
distributions for comparison with observations.
It has been shown in many recent papers that  hierarchical galaxy formation models
can reproduce many observed properties of galaxies both at low and at high redshifts.
Some highlights include variations in galaxy clustering with luminosity,
morphology and redshift 
(Kauffmann, Nusser \& Steinmetz 1997; Kauffmann et al 1999b, Baugh et al 1999), 
the evolution
of cluster galaxies (in particular of ellipticals)                                     
(Kauffmann 1995; Kauffmann \& Charlot 1998) and the properties of 
the Lyman break galaxy population at $z \sim 3$
(Baugh et al 1998; Governato et al. 1998; Somerville, Primack \& Faber 1999; Mo, Mao \&
White 1999).

In these models, the quiescent accretion of gas from the halo results in the formation of a disk.
If two galaxies of comparable mass merge, a spheroid is formed.
It has been demonstrated that a merger origin for ellipticals can explain    
both their detailed internal  structure
(see for example 
Barnes 1988; Hernquist 1992,1993; Hernquist, Spergel \& Heyl 1993; 
Heyl, Hernquist \& Spergel 1994) 
and global population properties 
such as  the slope and scatter of the colour-magnitude relation  
and its evolution to high redshift (Kauffmann \& Charlot 1998). 
Simulations including a gas component 
(Negroponte \& White 1983; Barnes \& Hernquist 1991, 1996; Mihos \& Hernquist 1994) 
have also shown  that mergers  drive gas far enough inwards to fuel 
nuclear starbursts, and probably  also central black holes. 
This is  the standard paradigm for the orgin 
of ultra-luminous infrared galaxies (ULIRGs), which 
can have star formation rates in excess of several hundred solar masses a year 
and are almost always associated with merging or interacting systems 
(Sanders \& Mirabel 1996). A significant fraction of the ULIRGs also
exhibit evidence of AGN activity (Genzel et al 1998).

In this paper, we assume that major mergers are responsible for the
growth and fuelling of black
holes in galactic nuclei. If two galaxies of comparable mass merge, the
central black holes of the progenitors coalesce and
a few percent of the gas in the merger remnant is
accreted by the new black hole on a timescale of a few times $10^7$ years.
Under these simple assumptions, the model is able to reproduce both 
the relation between bulge luminosity and black hole mass observed in nearby galaxies and
the evolution of the quasar luminosity function with redshift. 
Our models also 
fit many aspects of the observed evolution of galaxies,
including the present-day K-band luminosity function, the evolution of
the star formation rate density
as function of redshift, and the evolution of the total mass density 
in cold gas  inferred from observations of damped Lyman-alpha systems.
Finally, we study the evolution
of the abundance of starbursting systems and the relationship
between the luminosities of quasars and their host galaxies.

\section {Review of the Semi-Analytic Models of Galaxy Formation}

For most of this paper, the semi-analytic model we employ is that used by
Kauffmann \& Charlot (1998) to study the origin of the
colour-magnitude relation of elliptical galaxies formed by mergers
in a high-density ($\Omega=1$, $H_0$ = 50 km s$^{-1}$ Mpc$^{-1}$, $\sigma_8=0.67$) 
cold dark matter (CDM) Universe. The 
parameters we adopt  are those of Model ``A'' listed in Table
1 of that paper. In order to study how changes in  cosmological parameters affect
our results, we have also considered a low-density model with $\Omega=0.3$,
$\Lambda=0.7$, $H_0= 66$ km s$^{-1}$ Mpc$^{-1}$ , $\sigma_8=1$. This model is
discussed separately in section 7.

More details about semi-analytic techniques may be 
found in Kauffmann, White \& Guiderdoni (1993), Kauffmann et al. (1999a),
Cole et al. (1994) and Somerville \& Primack (1999). Below we present 
a brief summary of the main ingredients of the model. Because 
quasar evolution in our model  depends  strongly on the evolution of cold gas,
we pay particular attention to the parameters that control this. 
\begin {enumerate}
\item {\bf Merging history of dark matter halos.} We use an algorithm 
based on the extended Press-Schechter theory to generate Monte Carlo 
realizations of the merging paths of dark halos from high redshift
until the present (see Kauffmann \& White (1993) for details). 
This algorithm allows all the progenitors of a present-day object to
be traced back to arbitrarily early times.

\item {\bf The cooling, star formation and feedback cycle}
We have adopted the simple model for cooling first introduced by White
\& Frenk (1991). All the relevant cooling rate equations are described
in that paper. Dark matter halos are modelled as truncated isothermal 
spheres and it is assumed that as the halo forms, the gas relaxes to
a distribution that exactly parallels that of the dark matter. Gas
then cools, condenses and forms a rotationally supported disk at the 
centre of the halo. 

We adopt the {\em empirically-motivated} star formation law for disk 
galaxies suggested by Kennicutt (1998), which has the form
$ \dot{M}_* = \alpha M_{\rm cold}/ t_{\rm dyn}$, where  $\alpha$ is a free
parameter and $t_{\rm dyn}$ is the dynamical time of the galaxy.
If angular momentum is conserved, the cold gas becomes rotationally 
supported once it has collapsed by a factor of 10 on average,
so the dynamical time may be written $ t_{\rm dyn}= 0.1 R_{\rm vir}/
V_{\rm c}$, 
where $R_{\rm vir}$ and $V_{\rm c}$ are the virial radius and 
circular velocity of the surrounding dark halo. Note that according 
to the simple spherical collapse model,  the virial radius scales 
with circular velocity and with redshift as $R_{\rm vir} \propto 
V_{\rm c} (1+z)^{-3/2}$, so that $t_{\rm dyn}$  scales with redshift
as  $(1+z)^{-3/2}$ and is independent of $V_c$.

Once stars form from the gas, it is assumed that 
supernovae  can reheat some of the cold gas to the virial temperature of the
halo. The amount of cold gas lost to the halo in time $\Delta t$ 
can be estimated, using simple energy conservation arguments, as 
\begin {equation} \Delta M_{\rm reheat} = \epsilon \frac {4}{3} 
\frac {\dot{M}_* \eta_{SN} E_{SN}} {V_c^2} \Delta t, \end {equation}
where $\epsilon$ is an efficiency parameter, $E_{\rm SN} \sim 10^{51}$ erg
is the kinetic energy of the ejecta from each supernova, and
$\eta_{\rm SN}$ is the number of supernovae expected
per solar mass of stars formed. The parameters $\alpha$ and $\epsilon$
together control the fraction of baryons in the form of hot gas, cold 
gas and stars in dark matter  halos. In practice, adjusting $\epsilon$
changes the stellar mass of the galaxies, whereas adjusting  $\alpha$ 
changes their cold gas content. We choose $\epsilon$ to obtain a 
good fit to the observed present-day K-band galaxy luminosity function
and $\alpha$ to reproduce the present cold gas mass of our own Milky Way 
($\sim 4 \times 10^9 M_{\odot}$). We have also run 
models where $\alpha$ is not a constant, but varies with redshift as
$\alpha(z)= \alpha(0) (1+z)^{-\gamma}$. 
For positive $\gamma$,  less  gas is turned into stars per dynamical time 
in galaxies at high  redshift than at low redshift, as may be the case
if the star formation efficiency increases with time.
As shown in sections 3 and 5, this scaling of $\alpha$ 
has very little effect on the properties of galaxies at $z=0$,
but can produce  a much stronger evolution of the cold gas fractions to
high redshift and so a more strongly evolving quasar
population. In the following sections, we will show that $\gamma= 1-2$   
is required to fit both the evolution of the total mass density in cold gas
inferred from damped Ly$\alpha$ absorption sustems and
the rise in the space density of bright 
quasars from $z=0$ to $z=2$. 

\item {\bf Cooling flows in massive halos.}
As discussed in previous papers,  the cooling rates given by 
the White \& Frenk (1991) model lead to the formation of central 
cluster galaxies that are too bright and too blue to be consistent 
with observation if the cooling gas is assumed to form stars with a standard
initial mass function. Our usual ``fix'' for this problem has been to 
assume that gas cooling in halos with circular velocity greater than
some fixed value does not form visible stars. One question is whether
this material should then be  available to fuel a quasar. We
have considered two cases: 1) Gas that cools in halos with circular 
velocity greater than $600$ km s$^{-1}$ does not form visible stars, 
nor does it accrete  onto the central blackhole. 
2) Gas that cools in these massive halos does not form stars, but it 
is available to fuel quasars.

\item {\bf The merging of galaxies and the formation of ellipticals and bulges.}
As time proceeds, a halo will merge  with a number of
others, forming a new halo of larger mass. All gas that has not 
already cooled is assumed to be shock heated to the virial temperature
of this new halo. This hot gas then cools onto the central galaxy of the new
halo, which is identified with the central galaxy of its largest
progenitor. The central galaxies of the other progenitors become 
satellite galaxies, which are able to merge with the central galaxy 
on a dynamical friction timescale.
If two galaxies merge and the mass ratio between the satellite and the
central object is greater than 0.3, we add the stars of both objects 
together and create a bulge component. If $M_{\rm sat}/M_{\rm central} < 0.3$,
we add the stars and cold gas of the satellite to the disk component 
of the central galaxy. When a  bulge is formed by a merger, the cold 
gas not accreted by the blackhole is transformed into stars                                     
in a ``burst'' with a timescale of $10^8$ years. 
Further cooling of gas in the halo may lead to the formation of a new disk.
The morphological classification of galaxies is made according 
their B-band disk-to-bulge ratios. 
If $M(B)_{\rm bulge}-M(B)_{\rm total} < 1$ mag, then the galaxy is classified 
as early-type (elliptical or S0).

\item {\bf Stellar population models.} 
We use the new metallicity-dependent  stellar population synthesis
models of Bruzual \& Charlot  (1999,in preparation), which include 
updated stellar evolutionary tracks and new spectral libraries.
The chemical enrichment of galaxies is modelled as described in 
Kauffmann \& Charlot (1998). As chemical evolution plays  little role 
in our quasar models, we do not describe the recipes again in this paper.
\end {enumerate}

\section { The Global Evolution of Stars and Gas}

Figure 1 shows the K-band luminosity function of galaxies at $z=0$
compared with recent observational results. The K-band luminosities 
of galaxies are a good measure of their total stellar masses, rather
than their instantaneous star formation rates, and are affected very 
little by dust extinction. By normalizing in the K-band, we ensure
that our models have produced roughly the correct total mass density of stars
by the present day. As can be seen, virtually identical results are 
obtained for a model where the star formation efficiency $\alpha$ is 
constant and for a model where $\alpha \propto (1+z)^{-2}$. Figure 1 
also shows that cutting off star formation in cooling flows is
required in order to avoid producing too many very luminous galaxies.

\begin{figure}
\centerline{
\epsfxsize=8cm \epsfbox{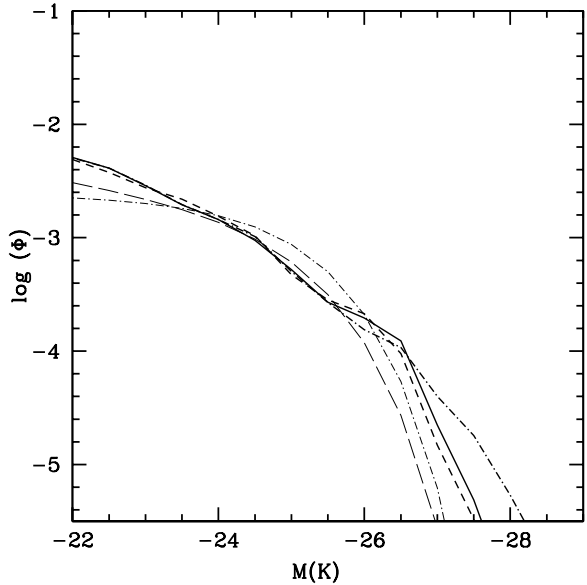}
}
\caption{\label{fig1}
\small
The present-day K-band luminosity function for galaxies. The thick 
dashed and solid lines are for the model with constant $\alpha$ 
and $\alpha \propto (1+z)^{-2}$ respectively. The dot-dashed line
show the result if stars {\em do} form in the cooling flows of massive
halos (see text). The thin lines are  Schechter fits to K-band 
luminosity functions derived by Gardner et al 1997 (dotted) and 
Szokoly et al 1998 (dashed)}                                     
\end {figure}
\normalsize

Figure 2 shows the evolution of the star formation rate density in the
models.  The data points plotted in figure 2 have been taken from
Lilly et al (1996), Connolly  et al (1997), Madau et al (1996) and 
Steidel et al (1999), and have
been corrected for the effects of dust extinction as described in
Steidel et al (1999). As can be seen, the evolution  of the SFR
density in the models agrees reasonably well with the
observations. Both show a factor $\sim 10$ increase from the present
day to $z \sim 1-2$, followed by a plateau. In the models, there is no
strong decrease in the SFR density until redshifts greater than 6. 
In the model where $\alpha$ evolves with redshift as $(1+z)^{-2}$, 
the percentage of stars formed in merger-induced bursts increases from
less tham 10\% at $z=0$ to 50\% at $z=2.5$. By redshift 4, two-thirds of the total
star formation  occurs in the burst mode. In the model with constant $\alpha$,
the fraction of stars formed in bursts increases much less, from $\sim$ 10 \% at
$z=0$ to  25 \% at high redshift. As will be demonstrated in the next sections,
the constant $\alpha$ model is unable to fit the observed increase in the quasar
space densities at high redshift.

\begin{figure}
\centerline{ \epsfxsize=8cm \epsfbox{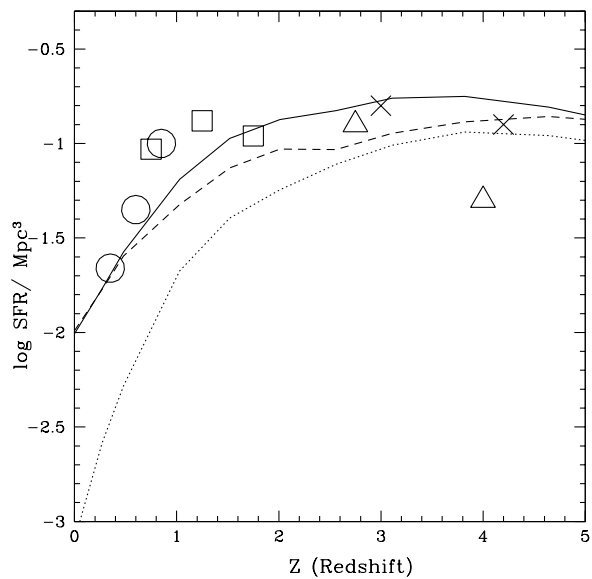} }
\caption{\label{fig2} \small The evolution of the star formation rate
density with redshift. The dashed and solid lines are for the
models with constant $\alpha$ and $\alpha \propto (1+z)^{-2}$
respectively. The dotted line shows the star formation rate density
occurring in merger-induced bursts in the model with $\alpha \propto (1+z)^{-2}$.
The open symbols  are  {\em extinction corrected} data points
read off from fig. 9 of Steidel et al (1999). These points are derived
from the data of Lilly et. al (1996) [circles], Connolly et al. (1997)
[squares], Madau et al. (1996) [triangles] and Steidel et al (1999)
[crosses].}
\end {figure}
\normalsize

Figure 3 compares the predicted evolution of the mean mass density in the
form of cold galactic gas in the models compared with  the values
derived from surveys of damped Ly$\alpha$ systems by Storrie-Lombardi,
McMahon \& Irwin (1996). As can be seen, the model with $\alpha
\propto (1+z)^{-1}$ agrees well with the data, but the model with
constant $\alpha$ severely underpredicts the  mass density of cold gas
at high redshifts. 
Note that the error bars
on the data points in figure 2 are large and that taking into account
the effects of dust  extinction would tend to move the points upwards
(Pei \& Fall 1995). The  model in which  $\alpha \propto (1+z)^{-2}$ 
 thus cannot be excluded. As we discuss in section 5, this model  leads to
the strongest evolution in the quasar space densities. 

\begin{figure}
\centerline{ \epsfxsize=8cm \epsfbox{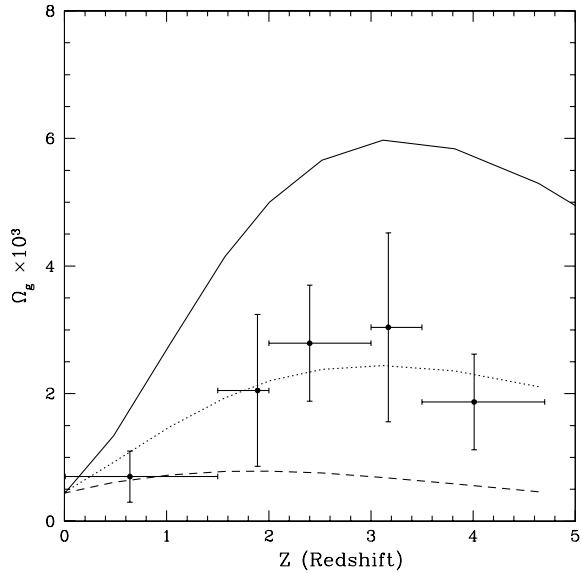} }
\caption{\label{fig3} \small The dashed, dotted and solid lines show the
cosmological mass density in cold gas in galaxies as a function of
redshift for the model with constant $\alpha$,
$\alpha \propto (1+z)^{-1}$ and $\alpha \propto
(1+z)^{-2}$ respectively.  The points with error bars show the  mean
cosmological mass density in neutral gas $\Omega_g$  contributed by
damped $Ly\alpha$ absorbers for $0.008 \le z \le 4.7$ from from
Storrie-Lombardi et al (1996).}
\end {figure}
\normalsize

\section{The Growth of Black Holes and the Bulge  Luminosity
--Black Hole  Mass Relation}                 

In our models, supermassive black holes grow by merging and  accretion
of gas during major mergers of galaxies. We assume that  when any merger
between two galaxies takes place,  the two pre-existing black holes in
the progenitor galaxies coalesce  instantaneously. In major mergers, some fraction of
the cold gas in the progenitor  galaxies is also accreted onto the new
black hole.  
As discussed in the previous section,  the cold gas fractions of galaxies increase strongly with
redshift. In hierarchical cosmologies, low mass bulges form                    
at higher redshift than high mass bulges.
To obtain the observed linear relation between black hole mass and bulge mass,
the fraction of  gas accreted by the black hole must be smaller for low mass galaxies,           
which  seems reasonable because gas is
more easily expelled from shallower  potential wells (equation 1).
We adopt a prescription in which the ratio of accreted mass to total
available cold gas mass scales with halo circular velocity in  the
same way as the mass of stars formed per unit mass of cooling gas.
For the parameters of our model, this may be written as 
\begin{equation}  M_{\rm acc} = \frac {f_{\rm BH} M_{\rm cold}}  
{1 + (280 \rm {km s} ^{-1} / V_c)^2}.
\end {equation} $f_{\rm BH}$ is a free parameter, which we set  
by matching to  the observed relation between bulge luminosity  and black
hole mass of Magorrian et al (1998) at a fiducial bulge luminosity $M_V= -19$. 
We obtain $f_{\rm BH} = 0.03$ for the model with $\alpha \propto (1+z)^{-2}$, 
$f_{\rm BH}=0.04$ for the model with
$\alpha \propto (1+z)^{-1}$ and 
$f_{\rm BH}=0.095$ for the constant  $\alpha$ model, a value which is
uncomfortably large.  

There are no doubt physical processes other than major mergers 
that contribute to the growth of supermassive black holes.
For example, we have neglected the accretion of gas during minor mergers
when a small satellite galaxy falls into a much larger galaxy (Hernquist \& Mihos 1995).  
We have also neglected the accretion of gas
from the surrounding hot halo (Fabian \& Rees 1995, Nulsen \& Fabian 1999).
It has been suggested that this may occur in the form of  advection dominated  accretion
flows (Narayan \& Yi 1995) and may produce the hard  
X-ray background (Di Matteo and Fabian 1997). If  such processes  
contribute significantly to the growth  of supermassive black holes,
we would need to lower the  fraction of the cold gas accreted by the
black hole during major mergers. 
The general trends predicted by our model would not be affected.

We have produced  absolute magnitude-limited catalogues of bulges
from our models and show scatterplots of black hole mass versus
bulge luminosity in figure 4. The thick solid  line
shows the relation derived by Magorrian et al and the dashed  lines
show the 1$\sigma$ scatter of their observational data 
around this relation. The first panel illustrates what happens  if we
add a fixed fraction of the gas to the black hole at each merging
event -- the relation is considerably too shallow. The  second  and
third panels show the relation obtained if the prescription in
equation 2 is adopted. As can be seen, the slope is now correct.   The
model with $\alpha \propto (1+z)^{-2}$ exhibits considerably more
scatter than the constant $\alpha$ model. This  scatter 
arises from the fact that bulges of given luminosity form 
over a wide range in redshift.  Because galaxies are more gas
rich at higher redshift, bulges that  form early will contain bigger
black holes than bulges that form late. The gas fractions of galaxies
rise more steeply in the model with $\alpha \propto (1+z)^{-2}$  than
in the model with $\alpha$ constant, giving rise to the increased scatter in panel 3.

One  observationally-testable prediction of our model is that
elliptical galaxies that formed recently should   harbour 
black holes with {\em smaller} masses than the spheroid 
population as a whole. This is
illustrated in the fourth panel of figure 4, where we show the
relation between bulge luminosity and black hole mass for ``isolated''
ellipticals in the $\alpha \propto (1+z)^{-2}$  model. 
These are elliptical  galaxies that reside
at the centres of dark matter halos of  intermediate mass; the fact
that they have not yet accreted a new disk component means that they
were formed by a major merger at most a few Gyr ago. To select such
objects in the real Universe, one would  look for ellipticals outside
clusters that have no neighbours of comparable or larger
luminosity within a radius of  $\sim 1.5 $ Mpc.
Conversely we predict that rich cluster ellipticals and bulges with large disks
should have relatively  massive black holes for their luminosity since these objects formed
early.

\begin{figure}
\centerline{ \epsfxsize=12cm \epsfbox{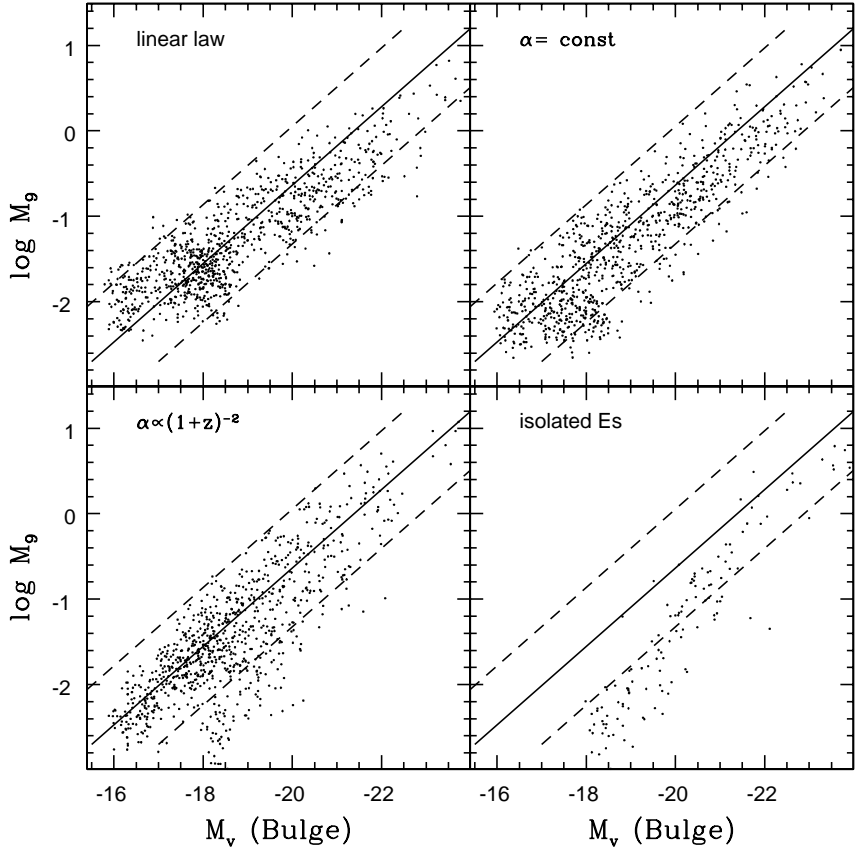} }
\caption{\label{fig4} \small The correlation between the 
logarithm of the mass of the central 
black hole expressed in units of $10^9 M_{\odot}$  
and the absolute V-band magnitude of the bulge. The dots are an absolute
 V-band magnitude limited sample of bulges in our model. The
thick solid line is the $M_V({\rm bulge})$ vs. $M_{{\rm BH}}$ relation obtained by
Magorrian et al (1998) for nearby normal galaxies. The dashed lines give an
indication of the 1$\sigma$ scatter in the observations.  The
difference between the four panels is explained in the text.}
\end {figure}
\normalsize

\section{ Evolution of the Quasar Luminosity Function}

Quasars have long been known to evolve very
strongly. Their comoving space density increases by
nearly two orders of magnitude from $z \sim 0.1$ to an apparent peak
at $z \sim 2.5$. The evolution at redshifts exceeding $2.5$ is
controversial: there is evidence that the number density of optically
bright quasars declines from $z \sim 2.5$ to $z \sim 5$
(Shaver et al. 1996, see Madau 1999 for a review)  , but it
remains to be seen whether the same is true for fainter objects or for
active galactic nuclei detected at X-ray wavelengths (Miyaji, Hasinger \& Schmidt 1998).

In a scenario where black holes grow by merging and by accretion of
gas, the number density of the most massive black holes increases
monotonically with time. This is illustrated in figure 5, where we
plot the black hole mass function in our model at a series of
redshifts. From now on, unless stated otherwise, we only show results
for our ``fiducial'' model in which the star formation efficiency
$\alpha$ scales as  $(1+z)^{-2}$ and gas cooling in halos with circular
velocities greater than 600 km s$^{-1}$ is not accreted by black holes.

\begin{figure}
\centerline{ \epsfxsize=8cm \epsfbox{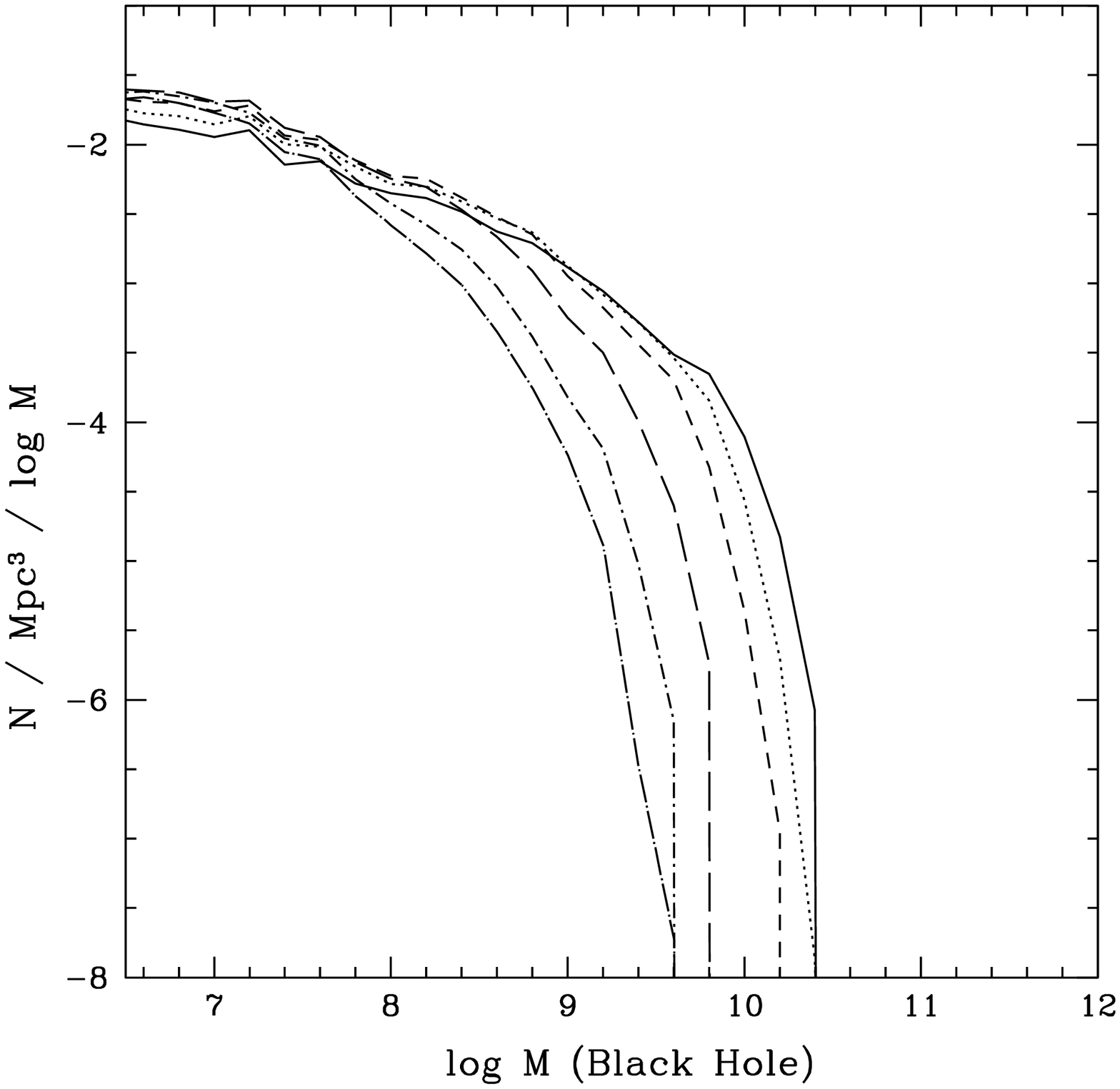} }
\caption{\label{fig5} \small The mass function of black holes as a
function of epoch. Solid, dotted, short-dashed, long-dashed, short
dashed-dotted and long dashed-dotted show results at z=0, 0.5, 1, 2,
3.1 and 3.8 respectively.}
\end {figure}
\normalsize

Quasars are activated when  two galaxies (and their central black
holes) merge and fresh gas is accreted onto the new black hole in the
remnant. Merging rates increase with redshift  in
hierarchical cosmologies and it has been suggested that this alone 
might explain the observed evolution of quasars (Carlberg 1990). Figure 6 demonstrates
that mergers alone will not do the job.  We plot the evolution of the
number density of the major mergers  that produce black holes of various
masses. The number density of merging events producing black holes of
$10^{10} M_{\odot}$ {\em decreases} at high redshift,  simply because 
such massive objects form very late.  The number density of mergers producing
smaller black holes does increase to high redshift, but the effect
is  too small to explain the observed increase in quasar space
densities from $z=0$ to $z=2$.

\begin{figure}
\centerline{ \epsfxsize=8cm \epsfbox{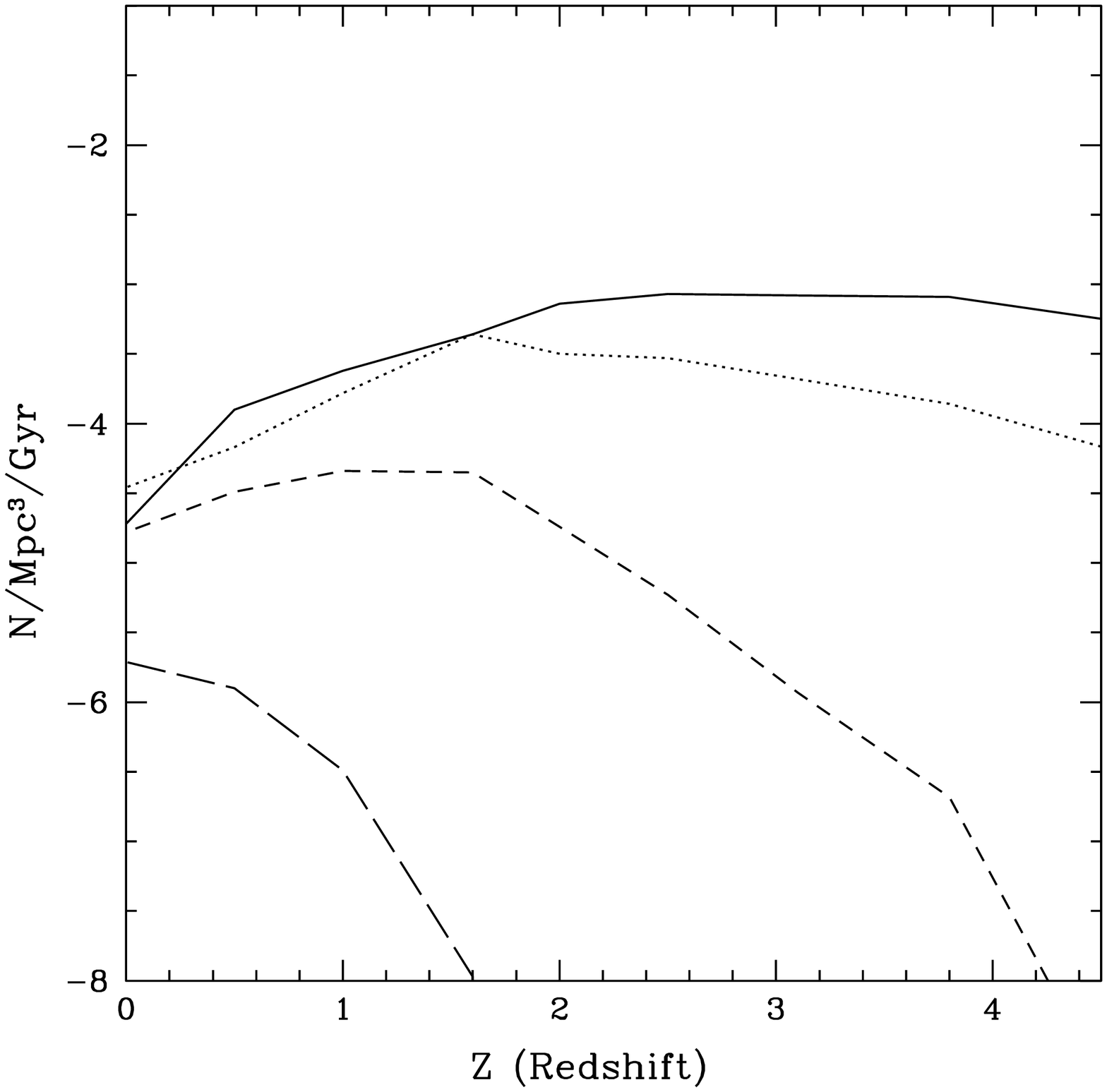} }
\caption{\label{fig6} \small The evolution of the number
of major mergers per Gigayear and per comoving cubic Mpc which produce
black holes of different mass. Solid, dotted, short-dashed and long-dashed
lines show results for merged galaxies containing black holes with $10^7$,
$10^8$, $10^9$ and $10^{10} M_{\odot}$ respectively.}
\end {figure}
\normalsize

Another hypothesis is that black holes run out of  fuel at
late times.  That such an effect does exist in our models is shown in
figure 7, where we plot  the redshift dependence of  the 
amount of gas accreted by merged  black holes of given mass.              
The amount of accreted gas accreted 
increases by a factor $\sim
3$ from $z=0$ to $z=1$. Note that in  
our models quasars do not run out of fuel because the gas supply is almost exhausted at present,
but because cool gas is converted into stars more efficiently at low redshift.

\begin{figure}
\centerline{ \epsfxsize=8cm \epsfbox{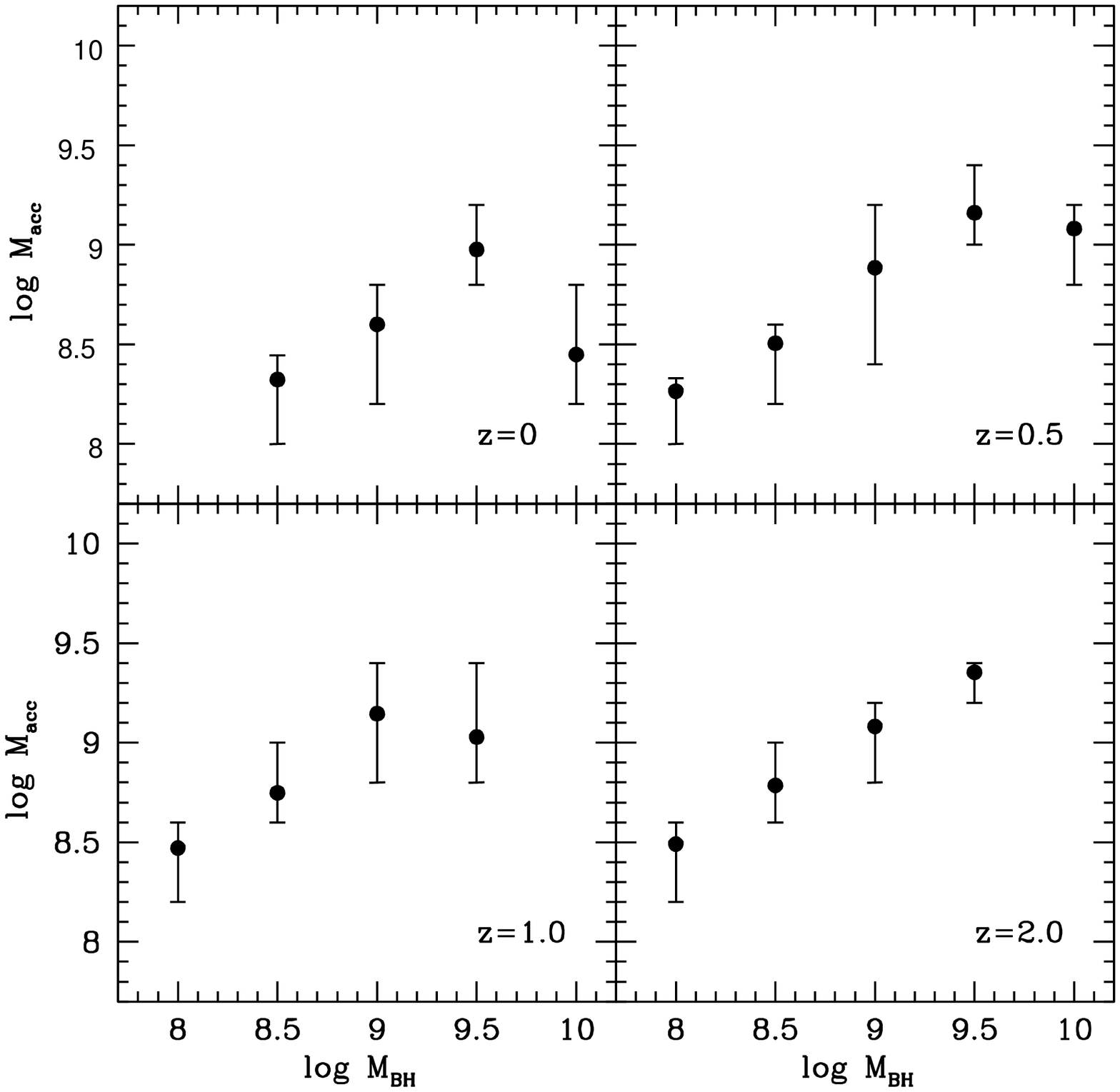} }
\caption{\label{fig7} \small The average mass of gas accreted by merged
black holes of mass $M_{\rm BH}$ after major merging events. The four panels refer to
four different
epochs. The error bars show the 25th and 75th percentiles of the
distribution.}
\end {figure}
\normalsize

In order to make direct comparisons with observational data, we must 
specify how a black hole accretion event turns into a quasar light 
curve.  We assume that a fixed fraction $\epsilon_B$ of the  rest
mass energy of the accreted material  is radiated in the B-band. This
results in the following transformation between  the accreted gas mass
$M_{\rm acc}$ and the absolute B-band magnitude of the quasar at the
peak of its light curve,
\begin {equation}  M_B ({\rm peak}) = -2.5 \log 
( \epsilon_B  M_{\rm acc}/t_{\rm acc}) -27.45. 
\end {equation} 
The timecscale $t_{\rm acc}$ over which gas is 
accreted onto the black hole during a merger 
is a second parameter of the model and
we have to make some assumption as to its scaling with mass and redshift.
We explore 
two possibilities:  1)$t_{\rm acc} \propto (1+z)^{-1.5}$. In this case, 
$t_{\rm acc}$ scales with redshift in
the same way as  $t_{\rm dyn}$, as expected if  the radius of the  accretion  
disk were to scale  with the radius of the  galaxy.   
2) $t_{\rm acc}$ = constant. In both cases we assume $t_{\rm acc}$ to be independent of mass.

Note that if we change the radiative efficiency parameter
$\epsilon_B$,  the quasar luminosities simply scale  by a constant
factor. If we change the accretion timescale $t_{\rm acc}$, we  affect
both the number densities and the luminosities of the quasars. 
It is usually assumed that the luminosities of quasars should  not  exceed the 
Eddington limit and we therefore introduce an upper limit to the 
B-band luminosity of
quasars that scales linearly with the mass of the black
hole $  L_{B} ({\rm max}) = 0.1 L_{\rm edd} \sim 0.14 (M_{\rm bh}/10^{8} M_{\odot}) 10^{46}$ erg s$^{-1} $.  
Some of  our quasars accrete at rates higher than that 
necessary to  sustain the Eddington luminosity, especially at high
redshift and in models with short $t_{\rm acc}$.
These  quasars  nevertheless obey the 
Eddington limit  because  a ``trapping surface'' develops within 
which the radiation advects inwards rather than escapes. 
As a result, the emission efficiency declines inversely with the 
accretion rate for such objects (Begelman 1978).

Finally, we assume that the luminosity of a quasar a time $t$ after the
merging event declines as
\begin {equation}   L_B(t) = L_B (\rm{peak}) \exp (-t/t_{\rm acc}).  
\end {equation}

We have chosen to normalize our model luminosity function to the
abundance of bright ($M_B < -24$) quasars at redshift 2. We have
chosen three values of $t_{\rm acc}$ for illustrative purposes: $10^7$,
$3 \times 10^7$, and    $10^8$ years.  
The typical  values derived  
for $\epsilon_B$ are in the range  $\sim 0.002-0.008$. This leaves 
some room for emission in  accretion modes other than that 
traced by optically bright quasars.  Two possibilities 
we do not treat here are advection dominated accretion flows 
and dust-obscured accretion (see Haehnelt, Natarajan \& Rees 1998
for a recent discussion).  

The resulting luminosity
functions  in four different redshift intervals are shown for 
our  fiducial model in figure 8.  
For this plot we assumed that the accretion time scales in the same way as the
host galaxy dynamical time, $t_{\rm acc} 
\propto (1+z)^{-1.5}$.  
The data points plotted as filled circles are taken from the
compilation by Hartwick \& Schade(1990). The triangles are from 
the Hamburg/ESO bright quasar survey by Koehler et al (1997). 
These authors find that luminous quasars are much more numerous in the
local Universe than previous smaller surveys indicated. We find that  
an accretion timescale of about  $10^{7}\yr $ yield results that are
in reasonable agreement with the data. 
Figure 9 
shows the evolution of the space density of 
quasars with luminosity $M_B < -26$ and $M_B < -24$ as a function of redshift. 
The symbols on the plot represent data points compiled from a number of sources (see
figure caption for details). Note that at B-band  magnitudes brighter than $-26$,
the quasar number densities decrease sharply. As a result, small changes
in  model parameters, such as the treatment of gas in cooling flows, can make quite
a large difference to our results (see figure 11 below). We therefore regard
the comparison at $M_B < -24$ as a more robust test of the model.
Our ``best fit'' $\Omega=1$ CDM model with $\alpha \propto (1+z)^{-2}$ reproduces the
observed decline in quasar space density from $z=2$ to the present reasonably well,
but the corresponding evolution in the mean mass density of cold gas 
is stronger than inferred from the damped Lyman-alpha systems (figure 3).
As we will show in section 7, the $\Lambda$CDM model provides a better overall fit
to the observations.

\begin{figure}
\centerline{ \epsfxsize=14cm \epsfbox{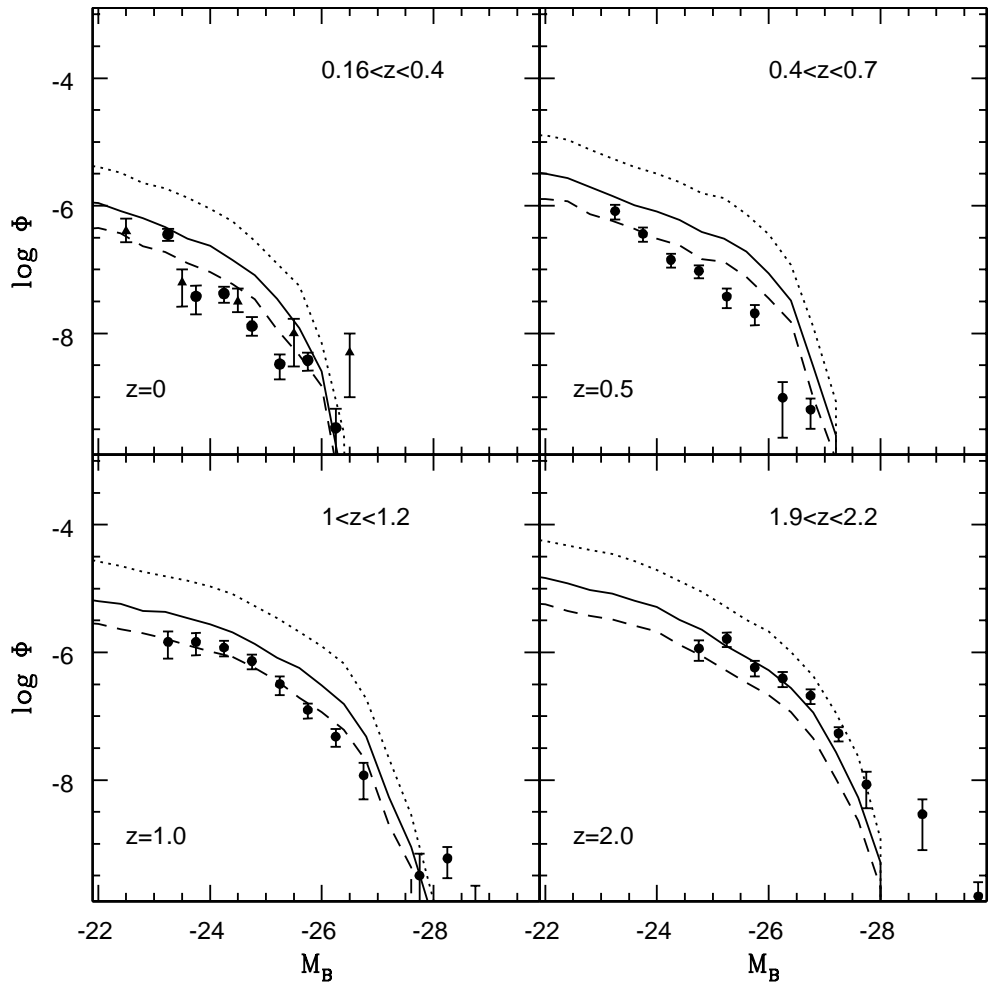} }
\caption{\label{fig8} \small The evolution of the B-band luminosity
function of quasars in our fiducial model. The solid lines are for
$t_{\rm acc}(0) = 3 \times 10^7$ years and the dashed and dotted  lines are for
$t_{\rm acc}(0)=10^7$ and $10^8$ years respectively.  The solid circles are data from the
compilation of Hartwick \& Schade (1990) and the triangles are taken from
Koehler et al (1997).}
\end {figure}
\normalsize

\begin{figure}
\centerline{ \epsfxsize=8cm \epsfbox{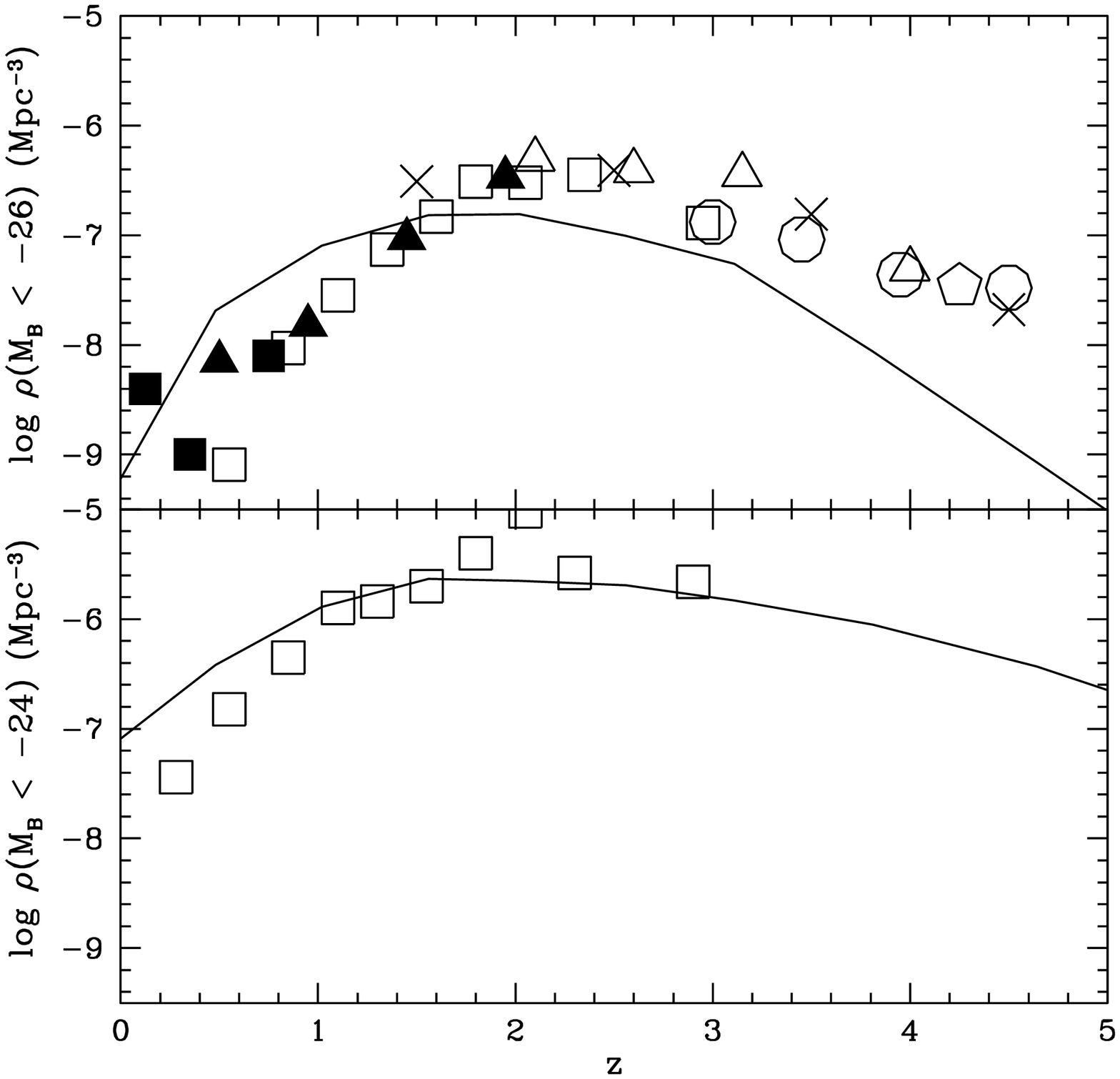}  }
\caption{\label{fig9} \small Top: the evolution of the space density of
quasars with $M_B < -26$ for  the model with $\alpha \propto (1+z)^{-2}$ and
$t_{\rm acc} = 10^7 (1+z)^{-1.5}$ Gyr (solid).  
Open squares show data from Hartwick \& Schade (1990), open triangles from Warren, Hewett \&
Osmer (1994), crosses from Hook, Shaver \& MacMahon (1998), 
open circles from Schmidt, Schneider \& Gunn (1995) and the open pentagon from 
Kennefick, Djorgovski \& De Carvalho (1995). The solid squares show data from 
Wisotzki et al (1998) and the solid triangles from Goldschmidt \& Miller (1998).
Bottom: the same for quasars with
$M_B < 24$. The data points are taken from Hartwick \& Schade (1990).} 
\end {figure}
\normalsize

In figure 10, we plot the mean ratio of quasar luminosity to Eddington
luminosity at a series of redshifts  for our best-fit model
with $t_{\rm acc}(z) = 1.0 \times 10^7 (1+z)^{-1.5}$ Gyr. The error
bars show the 25th and 75th percentiles of the distribution.
For faint quasars at low redshift, the  values of $L/L_{\rm Edd}$ range from 0.01 to 0.1.
These values increase for bright quasars and at high redshifts.
By $z=3$, $L/L_{\rm Edd}$ has increased to values  between 0.3 and 1. 
These results agree reasonably well 
with observed values of $L/L_{\rm Edd}$  inferred from the kinematics of the broad-line 
region,  X-ray variability and spectral  fitting of accretion 
disc models (see Wandel 1998 for a review; Laor 1998;  
Haiman \& Menou 1998; Salucci et al 1999).

\begin{figure}
\centerline{ \epsfxsize=10cm \epsfbox{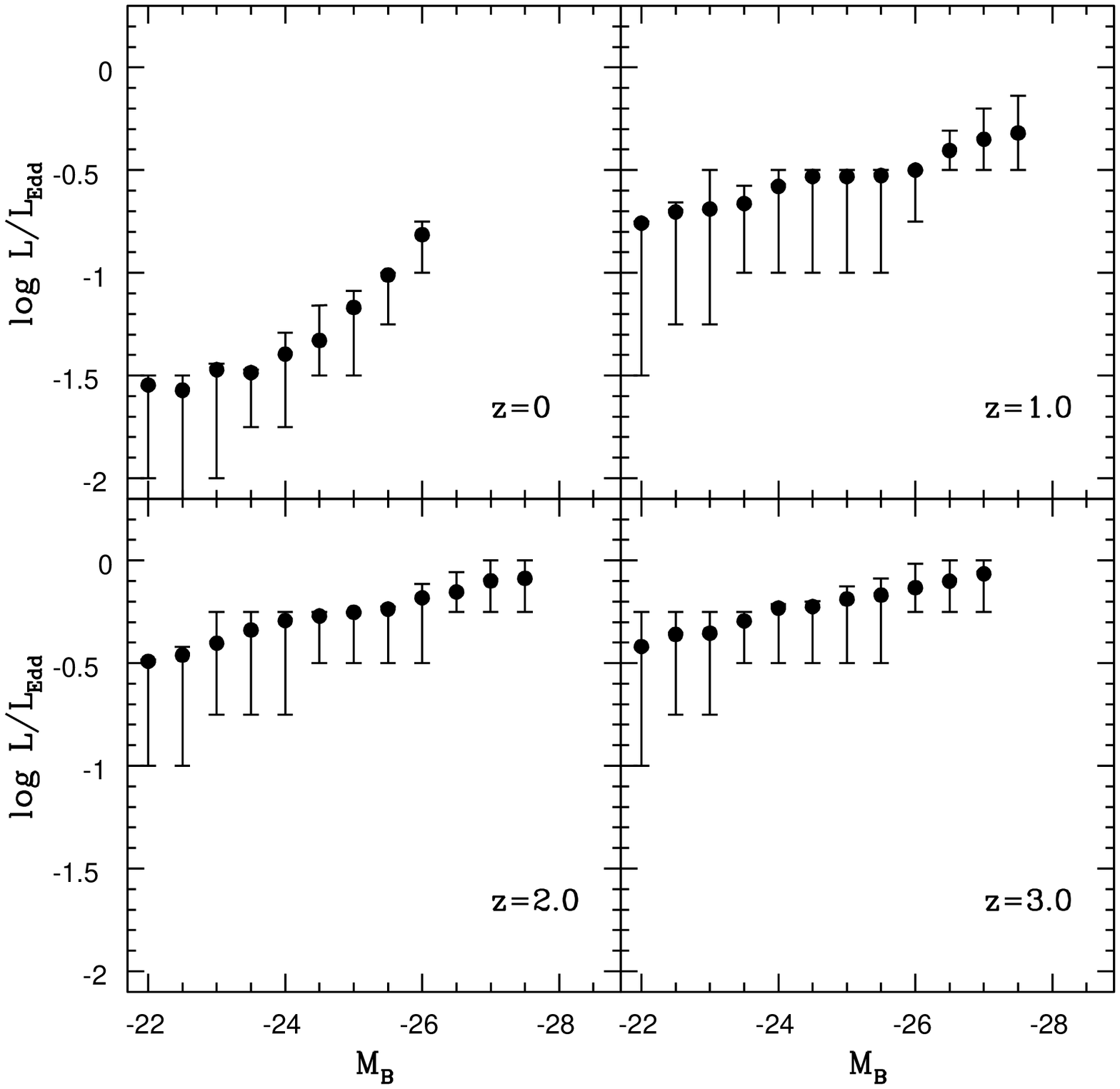} }
\caption{\label{fig10} \small The logarithm of the ratio of quasar luminosity to Eddington luminosity
is plotted as a function of the absolute B-band magnitude of the quasar for our
fiducial model with $t_{\rm acc}(z) =3.0 \times 10^7 (1+z)^{-1.5}$ Gyr. Error bars show the
25th to 75th percentiles of the distribution}
\end {figure}
\normalsize

We now study the sensitivity of our results to our
assumptions about cooling,  star formation and accretion timescales.
The leftmost panel in  figure 11 shows the evolution of the B-band
luminosity function for a model with $\alpha \propto (1+z)^{-1}$ and $t_{\rm acc} \propto
(1+z)^{-1.5}$.  In the other panels we show the effect of varying a
single parameter in the model:
\begin {enumerate}
\item {\bf Redshift dependence of the accretion time. } If $t_{\rm acc}$
is held contant, the space density  of  the brightest quasars evolves
very little from $z=0$ to $z=2$. 
\item {\bf Redshift dependence of $\alpha$}. If $\alpha$ is a
constant, the space density of quasars of all luminosities increases
much less from $z=0$ to $z=2$. As we have shown, a very strong scaling
of $\alpha$ with redshift ($\alpha \propto (1+z)^{-2}$)
is required in order to come reasonably close to fitting the
observed increase for an $\Omega=1$ CDM model.
\item {\bf  Cooling flow gas as fuel for black holes?} If gas in
cooling flows  can fuel  black holes, the number of very luminous
quasars at low redshifts increases substantially.  As a result,  the
evolution at the bright end of the luminosity function is weaker than
before. This model can fit the Koehler et al (1997) luminosity function
reasonably well at $z=0$.
The model also produces a luminosity function shape that is closer to 
a power-law at all redshifts.  
\end {enumerate}

\begin{figure}
\centerline{ \epsfxsize=14cm \epsfbox{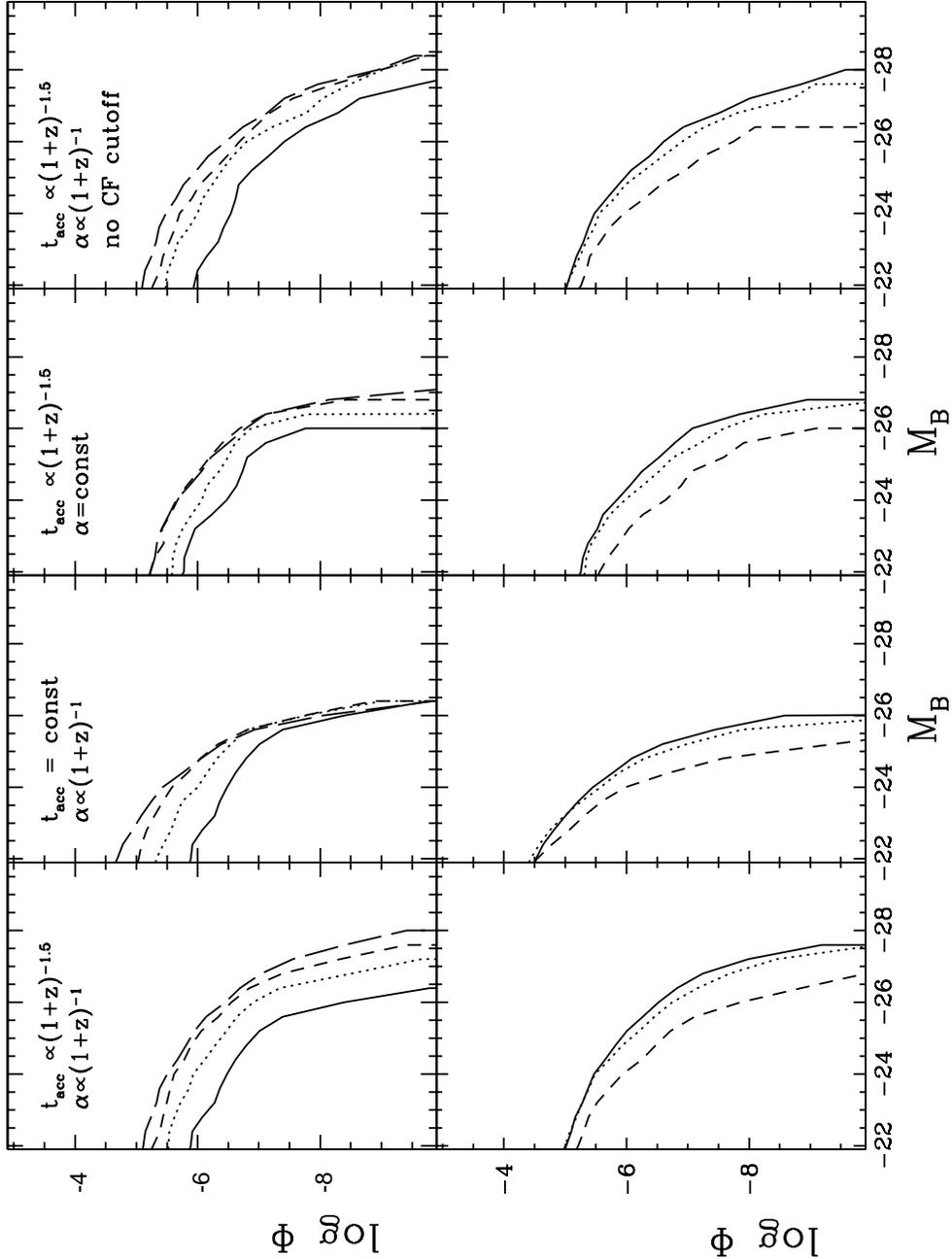} }
\caption{\label{fig11} \small A comparison of the B-band quasar
luminosity functions for (from left to right) a) a model with $t_{acc} \propto (1+z)^{-1.5}$
and $\alpha \propto (1+z)^{-1}$,
b) a model with $t_{\rm acc}=$ constant and $\alpha \propto (1+z)^{-1}$, c) a model with $\alpha$=
constant and $t_{acc} \propto (1+z)^{-1.5}$ and 
d) a model where gas in cooling flows is accreting onto
the black hole.  In the top panels, the luminosity functions are shown
at redshifts 0 (solid), 0.5 (dotted), 1 (short-dashed) and 2
(long-dashed). In the bottom panels,  The top panels, they are shown
at redshifts 2.5 (solid), 3.1 (dotted) and 4.6 (dashed).}
\end {figure}
\normalsize

\section {Further Model Predictions}

\subsection {The Host Galaxies of Quasars}
Recent near-infrared and Hubble Space Telescope (HST) imaging studies
of quasar host-galaxies at low redshifts show that luminous quasars
reside mainly in luminous early-type hosts (McLeod \& Rieke 1995;
Hutchings 1995; Taylor et al 1996; McLeod 1997; Bahcall et al 1997;
Boyce et al 1998; McLure et al 1998). There also appears to be an
upper bound to the quasar luminosity as a function of host galaxy
stellar mass (McLeod \& Rieke 1995).

In our models, quasars are only activated by major mergers which also
result in the formation of a  spheroidal remnant galaxy. By definition, all quasar
hosts are thus either ellipticals or spirals in the process of merging.
This is no doubt an oversimplified picture. It is
certainly possible    that minor mergers
or galaxy encounters trigger  gas accretion onto the
central black hole. It is also  likely that this is more important for
low luminosity quasars.

In figure 12, we show scatterplots of host galaxy luminosity versus quasar
luminosity at a series of different redshifts. For reference, the
horizontal line in each plot shows present-day value of $L_*$ for galaxies.
 At low redshift, quasars with magnitudes brighter than $M_B =
-23$ reside mostly in galaxies more luminous than
$L_*$. The luminosity of the host correlates with the luminosity of
the quasar, but there is substantial scatter ( typically a factor
$\sim 10$ in host galaxy luminosity at  fixed quasar B-band
magnitude).  At low quasar luminosities, the scatter is large because
the sample includes both low mass galaxies at peak quasar luminosity and high mass
galaxies seen some time  after the accretion event. At
high quasar luminosities, the scatter decreases and the sample
consists only of massive galaxies ``caught in the act''.  Our results
at low redshift agree remarkably well with a recent compilation of
ground-based and HST observations of quasar hosts  by
Mcleod, Rieke \& Storrie-Lombardi (1999).  In figure 12, we have drawn
a triangle around the region spanned by their observational data points.  At high
redshifts, our models predict that the quasars should be found in
progressively  {\em less luminous} host galaxies. This is not
surprising because in hierarchical models, the massive galaxies that
host luminous quasars at the present epoch are predicted to have
assembled recently (Kauffmann \& Charlot 1998). We caution, however,
that the the luminosities of quasars hosted by galaxies at different epochs depends strongly on
the redshift scaling of $t_{\rm acc}$.
As  discussed previously, smaller accretion timescales mean that
luminous quasars can be located in smaller host galaxies. In section 7, we explore
the extent to which the predicted masses of the host galaxies depend on the
choice of cosmology.

\begin{figure}
\centerline{ \epsfxsize=13cm \epsfbox{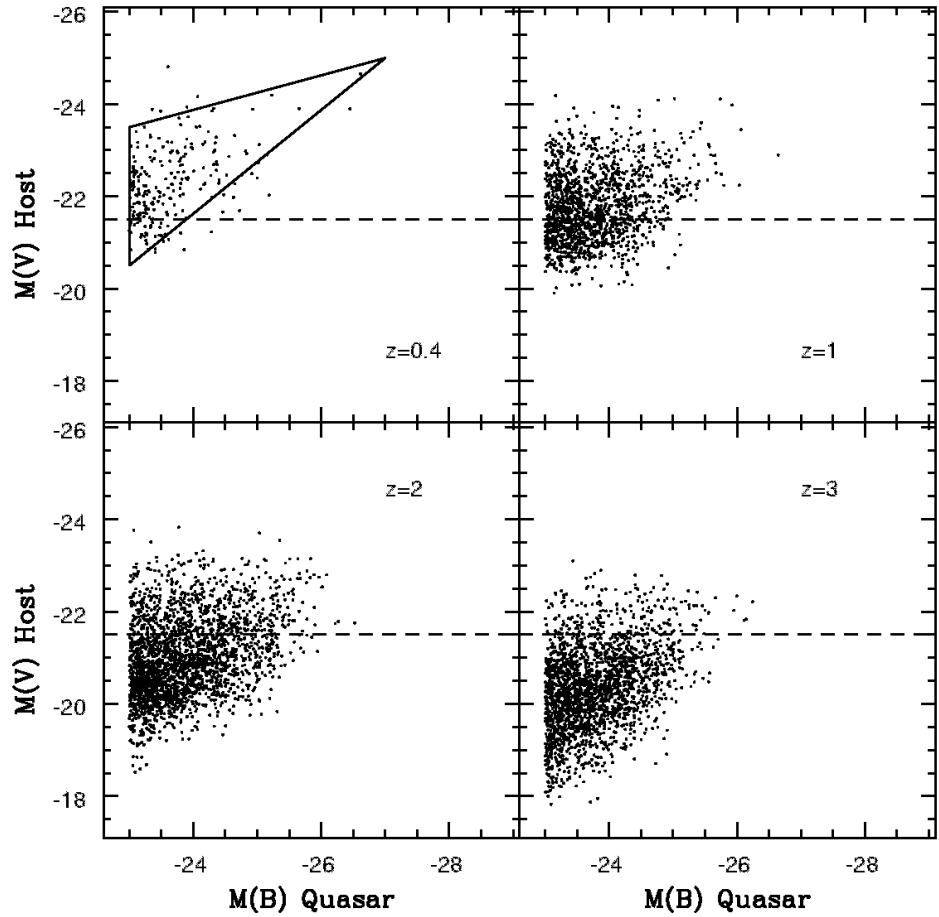} }
\caption{\label{fig12} \small Host galaxy versus quasar absolute
magnitudes at a series of redshifts. The dashed line shows the
present value of $L_*$ for galaxies. The triangular box in the top-left panel
shows the region spanned by the data set compiled by McLeod, Rieke \&
Storrie-Lombardi (1999).}
\end {figure}
\normalsize

\subsection {Evolution of Starbursting Galaxies} 
An important result from  the IRAS satellite  was the discovery
of galaxies with  luminosities in the far-infrared (8-1000 $\mu$m)
that exceed their optical or UV luminosities by factors of up to
80. The brightest ($> 10^{12} L_{\odot}$) of these objects
are often referred to as ultraluminous infrared galaxies (ULIRGs) and
their space densities are comparable to those of quasars of similar power.  The
nature of the energy source powering the ULIRGs has been a subject of
intense debate.  One hypothesis is that they are powered by
dust-embedded AGNs. Alternatively, their far-infrared luminosity may be
provided by an intense burst of star formation, with implied star
formation rates $\sim 100-1000 M_{\odot}$ yr$^{-1}$. Recently, Genzel
et al. (1998) have used ISO spectroscopy to argue that the majority
(70-80 \%) of these objects are dominated  by star
formation, with about $\sim 25\%$ powered by AGNs.  The Hubble Space
Telescope has been used to study the morphologies of a  sample of
150 ULIRGs (Borne et al 1999) and in almost all
cases,  the objects are made up of multiple subcomponents. This is
taken as evidence that most ULIRGs are interacting or merging systems.
 ULIRGs have now been detected at redshifts up to
$z\sim 1$ (Van der Werf et al 1999, Clements et al 1999), but the
total number of objects at high-z is currently too small
to draw firm conclusions about evolution rates.           
This situation will no doubt change with the next
generation of IR telescopes, SOFIA, SIRTF, IRIS and FIRST.

In figure 13, we show how the comoving number density of gas-rich
mergers evolves with redshift in the model with $\alpha \propto (1+z)^{-2}$. We plot the
number density  of mergers between galaxies containing more than
$10^9$, $10^{10}$ and $3 \times 10^{10}$ $M_{\odot}$ of cold gas
averaged over an interval of 1 Gyr. In order to transform the values
shown on the plot to a space density of starbursting galaxies, we
would need to make some assumptions about the typical timescale and
efficiency with which the gas is converted into stars, and whether or
not this is likely to scale with redshift.  Since there is not yet any
available data to which we can fit our predictions,  we prefer to
leave our plot in its ``raw'' form. For example  at $z=0$, if one
assumes a typical star formation timescale of $10^8$ years and a
gas-to-star conversion efficiency of 100\%, then one obtains a space
density of objects with star formation rates in excess of $100
M_{\odot}$ yr$^{-1}$ of $\sim 3 \times 10^{-6}$ Mpc$^{-3}$, in
reasonable agreement with what is observed.  Our models predict that
the space density of gas rich mergers evolves  strongly with redshift;
the density of mergers involving more than $10^{10}$ $M_{\odot}$ of
gas increases  by more than a factor of 50 from redshift 0, reaches a
very broad  peak at $z \sim 2-4$, and then declines again. Note that
this peak occurs at lower redshift for the most massive systems, simply because
they form later. 
Finally, our models predict that the ratio
of young stars formed in the  burst to that of ``old''
stars formed in the progenitors, should increase
strongly with redshift. This is illustrated in the bottom panel of
figure 13, where we show that  the ratio of the total mass of gas
$M_{\rm gas}$ to the total mass of stars $M_{\rm stars}$ in the two galaxies
before they merge evolves by a factor of $\sim 5-10$ from 0.1 at $z=0$ to 0.7
at $z=3$. In a $\Lambda$CDM model (section 7), there is a milder evolution
of the gas-to-star ratio with redshift.

\begin{figure}
\centerline{ \epsfxsize=8cm \epsfbox{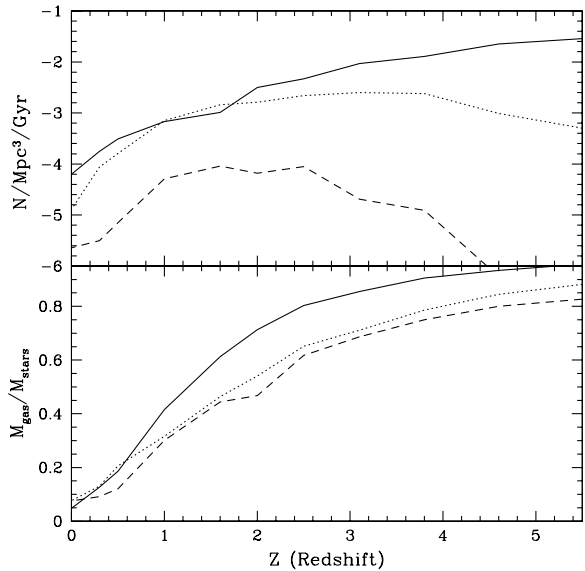} }
\caption{\label{fig13} \small Top: The evolution of the 
number of major mergers per Gyr and per comoving cubic Mpc  involving more than
$10^9$ (solid), $10^{10}$ (dotted) and $3 \times 10^{10}$ $M_{\odot}$
(dashed) of cold gas.  Bottom: The average ratio of gas mass to
stellar mass  during these mergers.}
\end {figure}
\normalsize

\subsection {Structural Properties of the Merger Remnants}
Low-luminosity ellipticals and spiral bulges
possess steeply rising central stellar density profiles that approximate power
laws, whereas high-luminosity ellipticals have central profiles with much shallower slopes
(termed cores). The power-law cusps in
small spheroids, their disky
isophotes and their rapid rotation, led Faber et al (1997) to suggest that they
formed in gas-rich dissipative mergers. 
Large spheroids, with their low rotation and boxy isophotes, plausibly
formed in near dissipationless mergers. In this case               
the orbital decay  of the massive black hole binary can            
 scour out a core in the stellar mass distribution
(Quinlan 1996; Quinlan \& Hernquist 1997).  One may ask whether
such a scenario is viable in the  hierarchical
merger models presented in this paper. Figure 14 illustrates that the
mergers that form low-luminosity ellipticals are indeed substantially more gas
rich than the mergers that form high-luminosity ellipticals. 
This is because small ellipticals typically form  at high
redshift  from mergers of low mass spirals with high gas fractions.
Conversely, massive ellipticals form a low redshift from massive spirals
that contain much less gas.
More detailed dynamical modelling is required in order to demonstrate that
this correlation is sufficient to explain   the observed dichotomy of
core profiles, isophote shapes and rotation speeds.

\begin{figure}
\centerline{ \epsfxsize=8cm \epsfbox{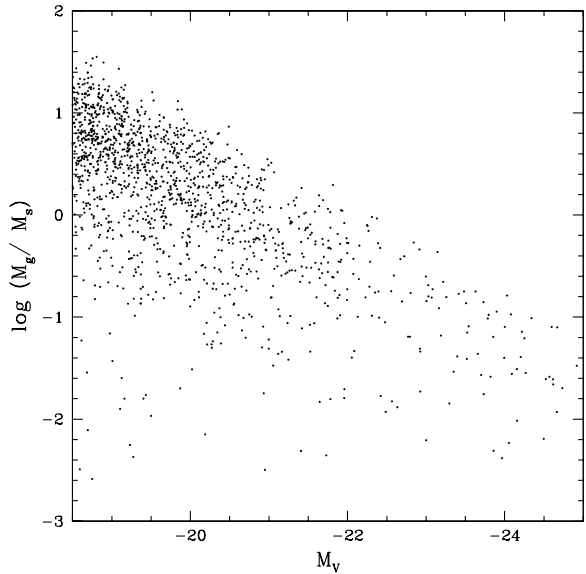} }
\caption{\label{fig14} \small The ratio of gas mass to stellar mass
present during the last major merger of an elliptical galaxy of
present-day magnitude $M_V$.}
\end {figure}
\normalsize

\section { The Influence of Cosmological Parameters}
All results shown so far have been for a CDM cosmology  with  $\Omega=1$, $H_0=$
50 km s$^{-1}$  and  $\sigma_8 =0.67$. Although we could explain the
evolution of the quasar luminosity function in a qualitative sense, the detailed        
fits to the observational data were not quite satisfactory. In particular, our     
``fiducial'' model in which the star formation efficiency parameter $\alpha$ scaled
with redshift as $\alpha \propto (1+z)^{-2}$ resulted in too much cold gas at
high redshift.

In figures 15-17 we show fits to the data obtained for  the popular $\Lambda$CDM
cosmology ($\Omega=0.3$,  $\Lambda=0.7$, $H_0=$ 70 km s$^{-1}$ and
$\sigma_8=1$). We note that quasar luminosity functions in the literature are almost always            
computed assuming a cosmology with $q_0=0.5$ and $H_0=$ 50 km s$^{-1}$.
In order to avoid having to re-analyze the real observational data, we choose to
transform our {\em model} predictions to the values that would be  obtained if the
observations were analyzed assuming an  Einstein-de Sitter universe with Hubble Constant $h=0.5$.
Because structure forms earlier in the $\Lambda$CDM cosmology, galaxy-galaxy merging rates
are substantially lower at $z=0$ than in the high-density CDM cosmology.
As a result, we do not require  an very strong evolution in the cold gas content of galaxies
to reproduce the strong decline in quasar space densities from  $z \sim 2$ to
the present day. Our best-fit model has $\alpha \propto (1+z)^{-1}$ and
$t_{acc} (z=0)= 2.5 \times 10^7$ Gyr. In addition we assume that $t_{\rm acc}$ scales in the
same way as the host galaxy dynamical time in a $\Lambda$CDM cosmology: $t_{\rm acc} \propto
(0.7 + 0.3(1+z)^3)^{-1/2}$ (Mo, Mao \& White 1998). As seen in figures 15-17,
the $\Lambda$CDM model can simultaneosly fit the evolution of cold gas inferred from the
damped systems {\em and} the decline in quasar space density from $z=2$ to the present day.

We also find that the masses of quasar host galaxies do not decline
as rapidly with redshift in this low-density model. As shown  in figure 18,
the host galaxies of  bright quasars in the SCDM model are slightly {\em more} massive than those 
in the $\Lambda$CDM at $z=0.5$. By $z=2$, however, the SCDM hosts are 60\% less massive
on average.

Most of the other observed properties of the galaxies and quasars in the best-fitting $\Lambda$CDM model, 
including the K-band luminosity function, 
the evolution
of the star formation rate density and the bulge luminosity/ black hole mass relation,
are  similar to those of the best-fitting SCDM model. For brevity, we will not show
these again.

\begin{figure}
\centerline{ \epsfxsize=7cm \epsfbox{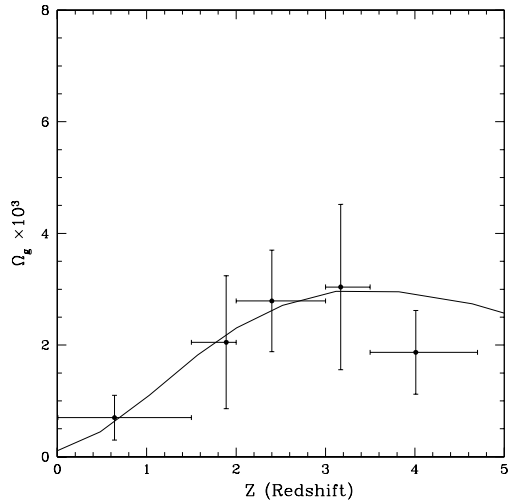} }
\caption{\label{fig15} \small The 
cosmological mass density in cold gas in galaxies as a function of
redshift for the $\Lambda$CDM model ($\alpha \propto (1+z)^{-1}$).}           
\end {figure}
\normalsize

\begin{figure}
\centerline{ \epsfxsize=10cm \epsfbox{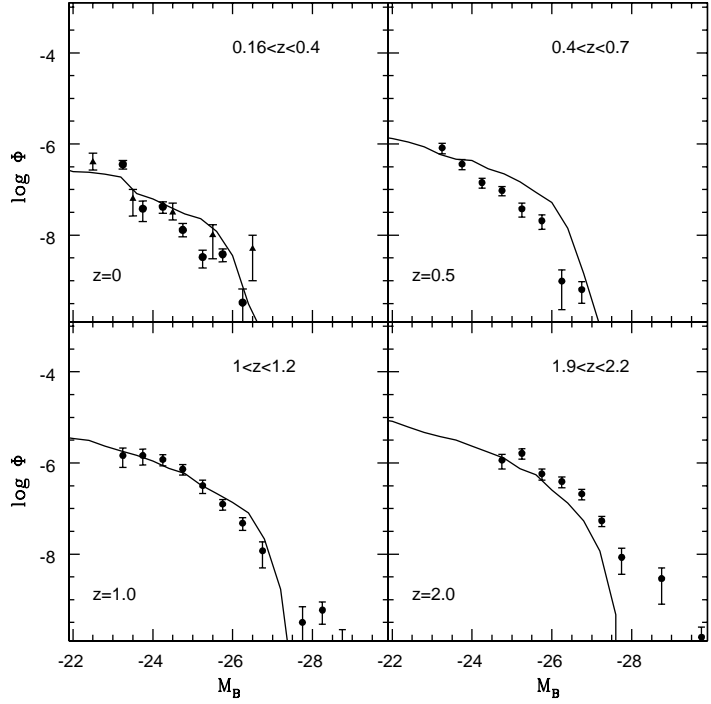} }
\caption{\label{fig16} \small The evolution of the B-band luminosity
function of quasars in the $\Lambda$CDM model. The model shown has    
$\alpha \propto (1+z)^{-1}$ and                        
$t_{\rm acc} = 2.5 \times 10^7 (0.7+0.3(1+z)^3)^{-1/2}$ Gyr. } 
\end {figure}
\normalsize

\begin{figure}
\centerline{ \epsfxsize=7cm \epsfbox{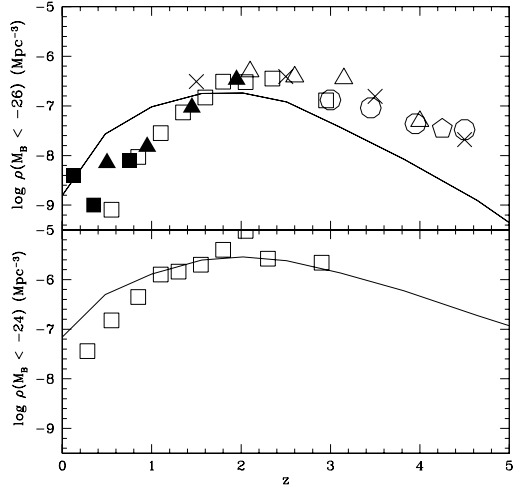}  }
\caption{\label{fig17} \small Top: the evolution of the space density of
quasars with $M_B < -26$ for  the $\Lambda$CDM model.  
Bottom: the same for quasars with
$M_B <24$. } 
\end {figure}
\normalsize

\begin{figure}
\centerline{ \epsfxsize=12cm \epsfbox{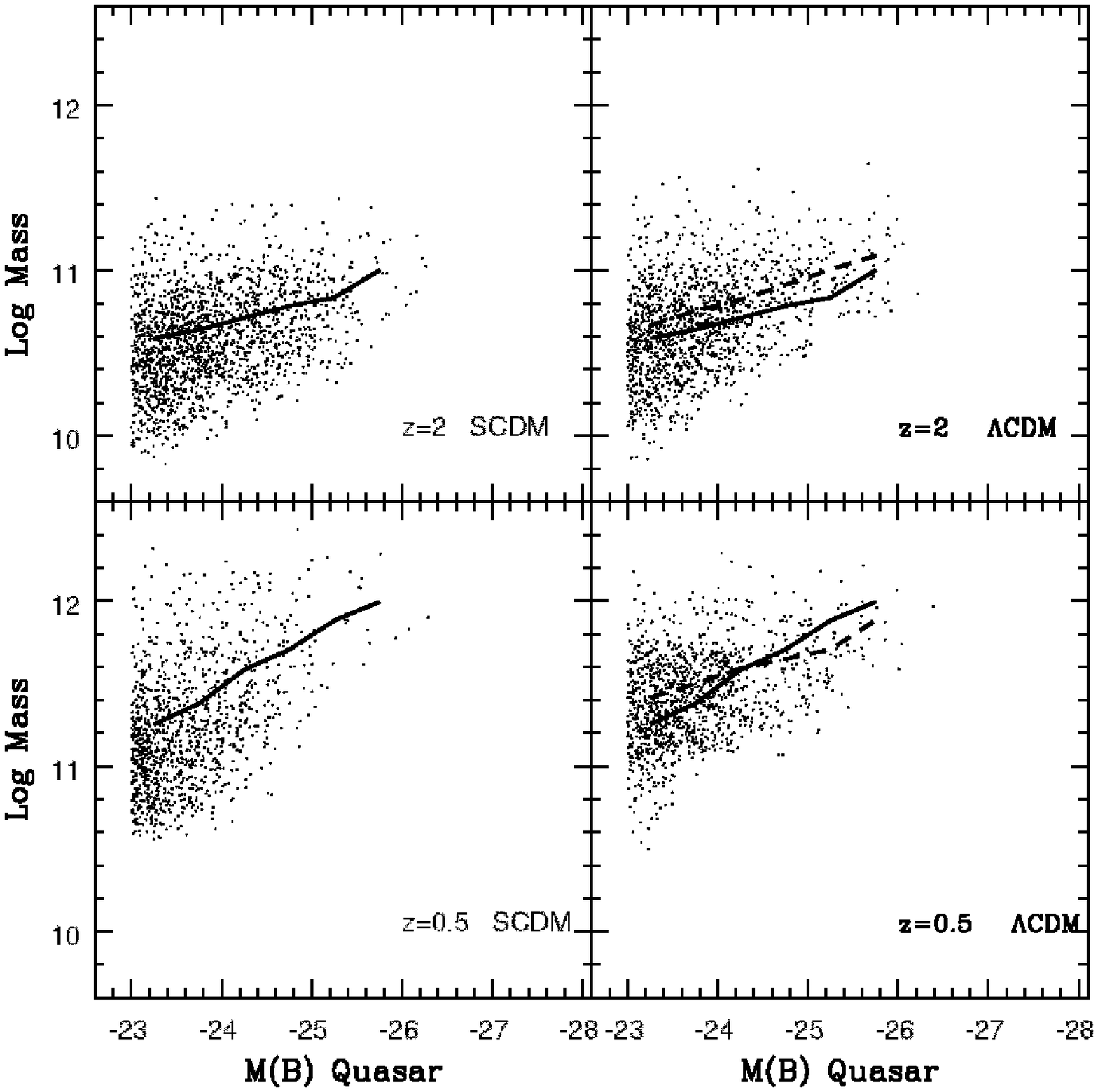} }
\caption{\label{fig18} \small Log of host galaxy mass versus quasar absolute
magnitudes for the SCDM and $\Lambda$CDM models . The thick solid line shows the
average host mass as a function of quasar magnitude in the SCDM model, while
the thick dashed line shows the same relation in the $\Lambda$CDM model.}
\end {figure}
\normalsize

\section {Summary \& Discussion}
 
The aim of this paper has been to demonstrate that the redshift evolution of galaxies
and quasars can be explained in a unified way within hierarchical models of structure formation.
This appears possible provided:            
a) black holes form  and grow mainly during the major mergers
that are responsible for the  formation of ellipticals;  
and b) the gas  consumption efficiency in galaxies scales with redshift
so that galaxies have higher cold gas fractions at earlier times.           

We have set the free parameters controlling star formation and feedback in our model
to match the stellar mass function of present-day galaxies, the observed redshift  evolution
of the star formation rate density,  and the evolution
of the total cold gas content of the Universe as inferred from observations of
damped Lyman alpha systems.  We have assumed  that  during
major mergers,  the black holes in the progenitor galaxies coalesce
and a few percent of the available cold gas is accreted by the new black hole.
The fraction of accreted gas was chosen to match the observed relation between
bulge mass and black hole mass  at the present day.
We obtain a linear relation if the fuelling of black holes
is less efficient in low mass galaxies.
The scatter in the model relation arises because bulges form over a wide range in redshift.

The greatest success of our model is its ability to explain
the strong decrease in the space density of
bright quasars from $z=2$ to $z=0$. We  assume that when a black
hole accretes gas,  about 1 percent of the rest mass energy of this
material is radiated  in the B-band. The strong decrease in quasar activity  
results from a combination of three factors i) a
decrease in the  merging rates of intermediate mass galaxies at late
times,  ii) a decrease in the gas available to fuel the most massive
black holes and iii) the assumption  that black holes accrete gas more
slowly at late times. The evolution  of merging rates is the most
secure feature of the model, since it is a simple consequence of the growth of structure in
the  dark matter component. The Press-Schechter-based
algorithms that we employ have been tested against N-body simulations
and have found to work reasonably well ( Kauffmann \& White 1993; Lacey \& Cole 1994 ). 
The evolution of the gas supply,
on the other hand,
is strongly dependent on the chosen parametrization of star formation and
feedback in the models.
It is encouraging that the $\Lambda$CDM model in particular can 
match {\em both} the observed evolution of
quasars {\em and} the increase in  cold gas  with redshift inferred
from observations of damped Lyman alpha absorbers. 
Future HI observations 
of galaxies at high redshift will yield more direct information on how the
gas-to-stellar mass ratios of  galaxies evolve with lookback time.
The increase in the  gas fractions of galaxies at high redshift also  leads to a strong
evolution in the space density of merger-induced  starbursts.        
Our results favour rather short accretion times onto the central black hole ( $\sim
10^7$ years).

Our model also reproduces the luminosities of the host
galaxies  of low-redshift quasars.                            
We predict that quasar hosts are on average a factor 10                  
less massive  at $z=2$ than at $z=0$  if the accretion timescale evolves
$\propto (1+z)^{-3/2}$ as in our fiducial $\Omega=1$ model. If the accretion timescale evolves 
less strongly, the hosts
will be brighter. Haehnelt,
Natarajan \& Rees (1998) have  suggested that future measurements of
the clustering strengths of high redshift quasars should 
determine whether they reside in low mass (and thus weakly clustered)
galaxies  or in high mass systems  which should cluster
more strongly. The low-density $\Lambda$CDM model predicts a smaller drop in quasar
host mass at high redshift: the hosts are on average a factor $\sim 5$ less massive
at $z=2$ than at $z=0$.

Finally, we have presented some results on the nature of the 
merging  process by which ellipticals form.  One  basic
feature of the hierarchical galaxy formation  scenario is that the most
massive spheroids  form {\em late}. The strong decrease
in the cold gas content of galaxies  inferred from damped Ly$\alpha$ systems and required to
explain the observed evolution of quasars means that the most massive ellipticals
must form in nearly dissipationless mergers.         
It is interesting that dissipationless
mergers may also be required to explain the core structure and the boxy isophotes observed in
massive ellipticals.

Our assumptions  regarding the growth of
supermassive black holes and the  resulting optical  quasar emission
are deliberately  simple.        
In reality these processes will probably  depend in a more
complicated way on the properties of the merging galaxies. Furthermore
there should be some  accretion  due to minor mergers and non-merging encounters, as well as   
accretion from  
the hot gas halos of elliptical galaxies and clusters. 
Nevertheless, our results demonstrate that 
the evolution of galaxies, the growth of
supermassive black  holes and the evolution of quasars and starbursts,
can all be explained in consistent way in a  hierarchical cosmogony. 

\vspace{0.8cm}

\large
{\bf Acknowledgments}\\
\normalsize
We thank Christian Kaiser, Philip Best, Simon White and Huub Rottgering for
helpful discussions and comments on the manuscript.

\end {document}